\documentclass[preprints,article,accept,moreauthors,pdftex]{Definitions/mdpi} 
\firstpage{1} 
\pubvolume{1}
\articlenumber{0}
\doinum{}
\pubyear{2022}
\datepublished{} 
\hreflink{}
\pdfoutput=1

\ifdefined\rec
\renewcommand{\rec}[1]{\frac{1}{#1}}
\else
\newcommand{\rec}[1]{\frac{1}{#1}}
\fi
\ifdefined\c
\renewcommand{\c}[1]{\mathcal{#1}}
\else
\newcommand{\c}[1]{\mathcal{#1}}
\fi
\ifdefined\bs
\renewcommand{\bs}[1]{\boldsymbol{#1}}
\else
\newcommand{\bs}[1]{\boldsymbol{#1}}
\fi

\Title{\vspace{50pt}Event-by-event investigation of the two-particle source function in heavy-ion collisions with EPOS}

\TitleCitation{Event-by-event investigation of the two-particle source function in heavy-ion collisions with EPOS}

\Author{D\'aniel Kincses $^{1,}$*\orcidA{}, Maria Stefaniak $^{2,3}$\orcidB{}, M\'at\'e Csan\'ad $^{1}$\orcidC{}}

\AuthorNames{Daniel Kincses, Maria Stefaniak and Mate Csanad}

\AuthorCitation{Kincses, D., Stefaniak, M., Csanad, M.}

\address{%
$^{1}$ \quad E\"otv\"os Lor\'and University, H-1117 Budapest, P\'azm\'any P. s. 1/A, Hungary\\
$^{2}$ \quad GSI Helmholtzzentrum für Schwerionenforschung, Planckstr. 1, 64291 Darmstadt, Germany,\\
$^{3}$ \quad Warsaw University of Technology, pl. Politechniki 1, 00-661 Warsaw, Poland}

\corres{Correspondence: kincses@ttk.elte.hu}

\abstract{Exploring the shape of the pair-source function for particles such as pions or kaons has been an important goal of heavy-ion physics, and substantial effort has been made in order to understand the underlying physics behind the experimental observations of non-Gaussian behavior. In experiments, since no direct measurement of the source function is possible, quantum-statistical momentum correlations are utilized to gain information about the space-time geometry of the particle emitting source. Event generators, such as EPOS, however, provide direct access to the freeze-out coordinates of final state particles, and thus the source function can be constructed and investigated. The EPOS model is a sophisticated hybrid model where the initial stage evolution of the system is governed by Parton-Based Gribov-Regge theory, and subsequently a hydrodynamic evolution is utilized, followed by hadronization and hadron dynamics. EPOS has already proven to be successful in describing several different experimental observations for systems characterized by baryon chemical potential close to zero, but so far the source shape has not been explored in detail. In this paper we discuss an event-by-event analysis of the two-particle source function in ${\sqrt{s_{NN}} = 200 \textrm{ GeV}}$ Au+Au collisions generated by the EPOS model. We find that when utilizing all stages of the model, L\'evy-shaped distributions (unlike Gaussian distributions) provide a good description of the source shape in the individual events. Hence it is clear that it is not the event averaging that creates the non-Gaussian features in the pair distributions. Based on this observation, we determine L\'evy-parameters of the source as a function of event centrality and particle momentum.}


\begin{document}
\section{Introduction}

A long-standing goal of high-energy nuclear physics has been to understand and explore the space-time geometry of the particle emitting source created in heavy-ion collisions~\cite{Lisa:2005dd}. One main observable that is of great interest is the two-particle source function, sometimes also called spatial correlation function or pair-separation distribution. Detailed investigation of this quantity is important for a multitude of reasons (connected to hydrodynamic expansion, critical behavior, etc.), however, it is not something that is easy to reconstruct experimentally. There is a whole sub-field of high-energy nuclear- and particle-physics called femtoscopy, which deals with such measurements of lengths and time intervals on the femtometer (fm) scale~\cite{Lednicky:2001qv}. Since it was shown by G. Goldhaber~et~al. that intensity correlations of identical pions can be used to gain information about the pair-source function~\cite{Goldhaber:1959mj,Goldhaber:1960sf}, femtoscopy has propelled to the forefront of investigations, and today it is still one of the most extensively studied field of high-energy physics. Besides the ample experimental studies, phenomenological investigations also placed great emphasis on describing the shape of the source function. Hydrodynamical model calculations suggest~\cite{Akkelin:1996sg, Akkelin:1995gh, Csorgo:1994fg, Csizmadia:1998ef, Csorgo:1995bi, Csanad:2009wc} that the source-shape is Gaussian, and this was adopted by many measurements as well~\cite{Adler:2004rq,Adams:2004yc}.

Source imaging studies~\cite{Afanasiev:2007kk,Adler:2006as} on the other hand suggest that the two-particle source function of pions has a long-range component, obeying a power-law behavior. It was also shown recently by various experimental measurements, that a generalization of the Gaussian source shape, the L\'evy distribution can provide a much more suitable description of the observed sources~\cite{Adare:2017vig,Porfy:2019rpi}. These kind of source shapes arise in many different scenarios~\cite{Csorgo:2003uv} such as anomalous diffusion~\cite{Csanad:2007fr}, jet fragmentation~\cite{Csorgo:2004sr}, critical behavior~\cite{Csorgo:2005it}, or resonance decays. It has been shown that even averaging over many events with different sizes can contribute to the appearance of a power-law component~\cite{Cimerman:2019hva,Cimerman:2020tpd}. In order to have a better understanding of the underlying processes behind the experimental results, more effort is needed from the phenomenology side. Among the important tools for such investigations are the event-generators that encompass different theoretical and phenomenological methods to model nuclear reactions. One of such event generators is the EPOS model~\cite{Werner:2010aa} -- the \textbf{E}nergy conserving quantum mechanical multiple scattering approach, based on \textbf{P}artons (parton ladders), \textbf{O}ff-shell remnants, and \textbf{S}plitting of parton ladders. In this paper we present a detailed event-by-event analysis of the two-pion source distribution in $\sqrt{s_{NN}} = 200$ GeV Au+Au collisions generated by EPOS. The event-by-event nature of our analysis helps in deciding if the role of event averaging is crucial in the apparent non-Gaussian but L\'evy nature of the observed sources.

The paper is structured as follows. In Section \ref{s:epos} we introduce the EPOS model and its several stages of evolution. In Section \ref{s:sourcefunction} we discuss the basic definitions and properties of the pair source distribution. In Section \ref{s:analysis} we discuss the details of the event-by-event analysis and the applied methods. In Section \ref{s:results} we present the results of the analysis and discuss our findings. Finally, in Section \ref{s:summary} we summarize and conclude.

\section{The EPOS model}
\label{s:epos}
EPOS, \textbf{E}nergy conserving quantum mechanical multiple scattering approach, based on \textbf{P}artons (parton ladders), \textbf{O}ff-shell remnants, and \textbf{S}aturation of parton ladders, is a phenomenological model based on Monte Carlo techniques. It opens up the possibility of investigating various phenomena and observables such as particle production, momentum distributions or flow correlations, providing a better understanding of the evolution of the system created in elementary (proton-proton) collisions and also during complex reactions involving heavy-ions. The theoretical framework included in the model provides a coherent description of the space-time expansion of matter based on a precise spectrum of studies of both elementary processes such as electron-positron annihilation or lepton-nucleon scattering and more compound collisions of protons or nuclei. The model was designed to describe processes appearing in collisions at $\mu_B \approx 0$, at very high (top RHIC or LHC) energies and for various systems, such as Au+Au, Pb+Pb or p+p. 

The EPOS model consists of several phases of evolution: 
\begin{itemize}
    \item initial stage (based on the Parton Gribov-Regge theory), 
    \item core/corona division,
    \item hydrodynamical evolution,
    \item hadronization,
    \item hadron rescattering,
    \item resonance decays.
\end{itemize} 
In this Section, all of these stages are described.

\subsection{Initial stage of the evolution}

In the theoretical framework of the model the crucial element is the sophisticated treatment of both the hadron-hadron scattering and the initial stage of the collisions at ultra-relativistic energies. It is highly relevant in the understanding of possible parton-hadron phase transitions. In EPOS, a merged approach of the Gribov-Regge Theory (GRT) and the eikonalised parton model is utilised to provide proper treatment of the first interactions happening just after a collision. This approach satisfies conservation laws, and treats the subsequent Pomerons (interactions) equally (as opposed to other multiple interaction approaches e.g. Pythia, where the first interaction is not treated exactly the same way as the others)~\cite{Drescher:2000ha}. 

\begin{figure}
\centering
\includegraphics[scale=0.5]{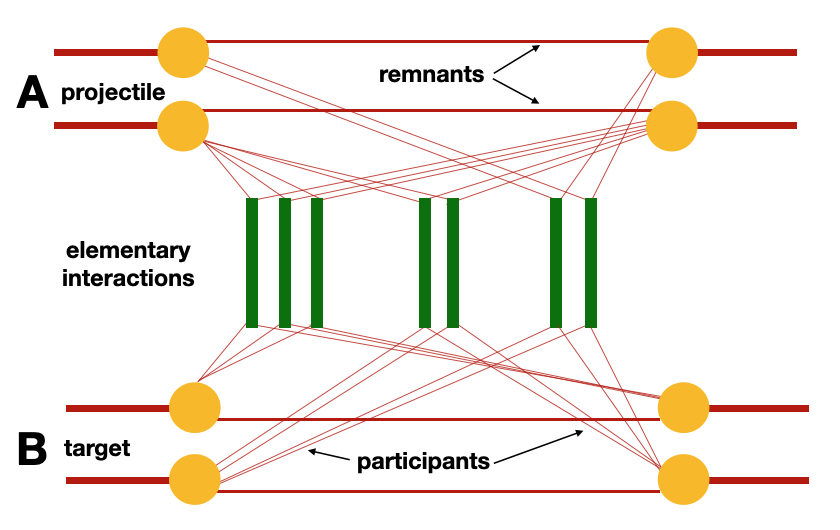}
\caption{An illustration of the nucleon-nucleon rescattering with two projectiles A and two targets B. The splitting between participants and remnants shows the momentum sharing between constituents ensuring conservation of the given variable~\cite{MS_thesis}.}
\label{fig:remnant}
\end{figure}

The formalism used for the calculation of the cross-sections is the same as the one used for calculating particle production. It is based on the Feynman diagrams of the QCD-inspired effective field theory and provides energy conservation. The nucleons are divided into a certain number of ``constituents'' carrying the incident momentum fraction. The fractions sum to unity in order to ensure momentum conservation. A nucleon is called a \textit{spectator} if it is not part of the interaction region of the colliding nuclei. If a nucleon is not a spectator, then its constituents can either be \textit{participants} taking part in the elementary interactions with constituents from the opposite side, or a \textit{remnant}, which although part of the interaction region, does not take part in the elementary interactions. This is illustrated on Fig. \ref{fig:remnant}.

\begin{figure}[h]
\centering
\includegraphics[scale=0.4]{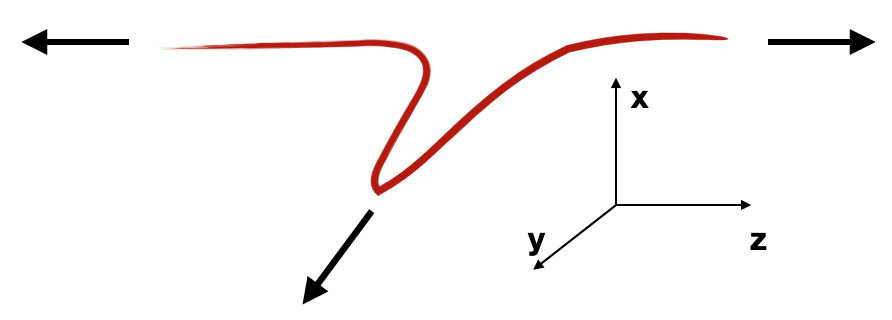}
\caption{The flux tube with the kink. The flux tube is mainly longitudinal but the kink part of the string moves transversely, here in the \textit{y}-direction~\cite{MS_thesis}.}
\label{fig:kink_ft}
\end{figure}

The particle production is based on the String Model approach \cite{Andersson:1986au, Ferreres-Sole:2018vgo}. The parton ladders are recognized as a quasi-longitudinal color field (elementary flux tubes) and are treated as classical strings \cite{Werner:2010aa}.
The intermediate gluons introduce the transverse motion into the \textit{kinky string} evolution. The schematic picture of the flux tube with the transverse kink is shown on Fig. \ref{fig:kink_ft}.

\subsection{Core-Corona approach}
If the density of the strings is very high, they cannot decay independently. This situation is characteristic for heavy-ion collisions and high-multiplicity $pp$ collisions. Henceforth in EPOS a dynamical process of the division of the strings segments into \textit{core} and \textit{corona} is introduced \cite{Werner:2013tya, Werner:2010aa, Werner:2007bf}.

The core-corona division is based on the abilities of a given string segment to leave the ``bulk matter''. The transverse momentum of the element and the local string density are considered as criteria for the division. If the string segment belongs to a very dense area, it will not escape but will contribute to the \textit{core}, which will be governed in the next step by the hydrodynamical evolution. When the segment originates from the part of the string close to the kink, it is characterized by high transverse momenta. It escapes the bulk matter and will contribute to the \textit{corona}. It consequently will show up as a hadron in a jet. There is also a possibility that the string segment is close to the surface of the dense part of the medium, and its momentum is high enough to leave it; then it also becomes a \textit{corona} particle. 

\subsection{Viscous Hydrodynamical evolution, event-by-event treatment and EoS}

In EPOS a 3D+1 viscous hydrodynamics approach is applied, called vHLLE (viscous relativistic Harten-Lax-van Leer-Einfeld Riemann solver-based algorithm). In the simulations, the separate treatment of individual events is highly important -- smooth initial conditions for all events are not applied. The event-by-event approach in hydrodynamical evolution is based on the random flux tube initial conditions \cite{Werner:2010aa}. It has a relevant impact on the final observables such as spectra or various harmonics of flow. The viscous hydrodynamics uses Equation of State X3F (``cross-over'' and ``3 flavor conservation'') which is compatible with lattice QCD data from Ref. \cite{Borsanyi:2010cj}. It corresponds to \mbox{$\mu_{B} = 0$ MeV}, hence this feature limits the applicability of the model to describe the region of the QCD phase diagram characterized by finite baryon density~\cite{Werner:2010aa}.

\subsection{Hadronization and Hadronic Cascades}

The expanding medium in the processes of hydrodynamical evolution reaching the given freeze-out condition is transformed into the particle spectra. In EPOS 3, the criterion characterizing the hadronization hypersurface is that the energy density equals \mbox{0.57 GeV/fm$^{3}$}. EPOS 3 furthermore utilizes the Cooper-Frye formula \cite{Cooper:1974mv} when determining distributions.
The final part of the simulation uses a so-called \textit{hadronic afterburner}, based on UrQMD \cite{Bleicher:1999xi,Bass:1998ca}. The hadronic scattering has a significant impact on the final observables \cite{Stefaniak:2018wwh}.

\section{The two-particle source function}
\label{s:sourcefunction}
\begin{figure}
    \centering
    \includegraphics[width=0.4\textwidth]{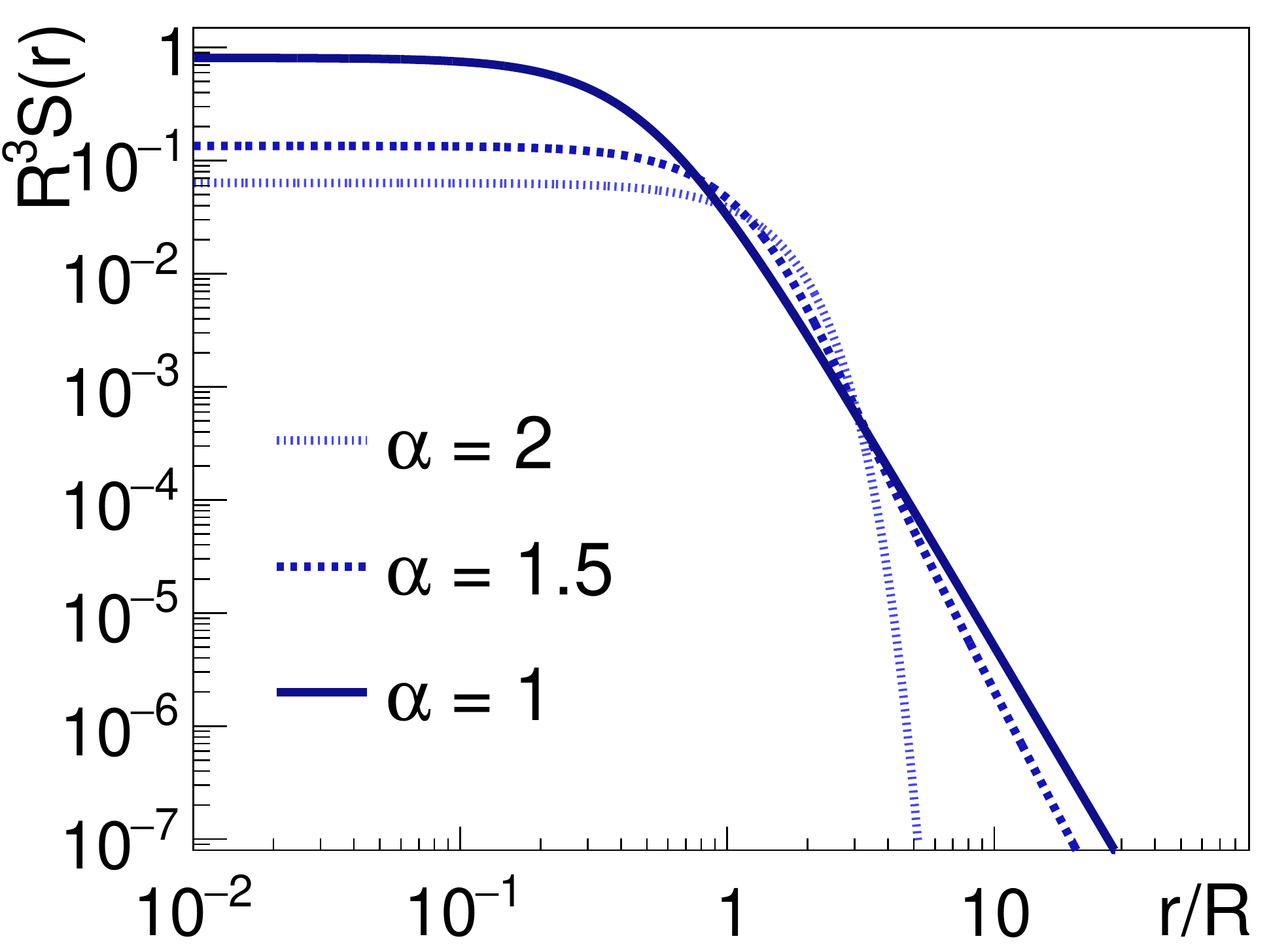}
    \caption{L\'evy-stable source distributions with $S(r) = \c L(|r|; \alpha, R)$ for $\alpha$ = 2, 1.5, and 1. The dependence on $R$ is scaled out.}
    \label{f:levyexample}
\end{figure}

In this section we discuss the basics definitions and properties of the two-particle source function. The pair source distribution is defined as the auto-correlation of the single particle phase-space density $S(x,p)$:
\begin{equation}
D(r,K) = \int S(x_1,K)S(x_2,K)d^4\rho = \int S(\rho + r/2,K)S(\rho-r/2,K)d^4\rho,
\label{eq:drk}
\end{equation}
where instead of the single particle variables $x$ and $p$ the pair-variables appear -- the pair center of mass four-vector $\rho=(x_1+x_2)/2$, the pair separation four-vector $r=x_1-x_2$, and the average momentum $K = (p_1 + p_2)/2$. The $D(r,K)$ distribution is the quantity that can be reconstructed indirectly from femtoscopic momentum correlation measurements, and experiments usually investigate the source-parameters that describe the shape of this distribution, see details e.g. in~\cite{Adare:2017vig}. It was recently shown by different experiments that for pions this pair-source exhibits a power-law behavior, and can be described with a L\'evy-stable distribution~\cite{Adare:2017vig,Porfy:2019rpi}. In case of spherical symmetry, the symmetric, centered stable distribution is defined as
\begin{equation}
\c L(\bs r; \alpha, R) =\frac{1}{(2\pi)^3} \int d^3\bs q e^{i\bs q\bs r}e^{-\rec{2}|\bs qR|^\alpha},
\label{eq:fx}
\end{equation}
where the temporal dimension is removed from the dependence, made possible by the mass-shell condition, as detailed in Ref.~\cite{Adare:2017vig}. The two important parameters that describe such a distribution are the L\'evy-scale parameter $R$ and the L\'evy-exponent $\alpha$. One can think of the latter as the parameter that is responsible for ``how far'' the distribution is from the Gaussian. In the $\alpha = 2$ case $\c L(\bs r; \alpha, R)$ is identical to a Gaussian distribution, while in case of $\alpha < 2$ it exhibits a power-law behavior. An illustration of the shape of such distributions can be seen on Fig.~\ref{f:levyexample}. Since this distribution retains the same $\alpha$ exponent under convolution of random variables, if the single-particle source densities have a L\'evy-shape then it follows that the two-particle source will also have such a shape, only the scale-parameter will be different. This can be summarized via the three-dimensional single-particle source $S(\bs x)$ and the pair-source $D(\bs r)$ (where we now suppress the momentum dependence, which is in turn contained in the parameters of the distribution):
\begin{equation}
S(\bs x) = \c L(\bs x; \alpha, R) \Rightarrow D(\bs r) = \c L(\bs r; \alpha, 2^{1/\alpha} R).
\end{equation}

There are already multiple experimental measurements for the L\'evy source parameters. Most notably the PHENIX experiment published results~\cite{Adare:2017vig} for 0-30\% centrality, $\sqrt{s_{NN}} = 200$ GeV Au+Au collisions, in the $m_T$ region of 0.23 GeV/$c^2$ to 0.87 GeV/$c^2$. They found the $\alpha$ parameter to be very slightly dependent on $m_T$, with an average value around 1.2. The NA61/SHINE experiment also has measured the L\'evy-exponent in 150A GeV Be+Be collisions, and obtained average $\alpha$ values of around 1.2~\cite{Porfy:2019rpi}.

\section{Analysis method}
\label{s:analysis}
For the analysis presented below we used $\sqrt{s_{_{NN}}}$ = 200 GeV Au+Au events generated by EPOS359. Using like-sign pion pairs, we measured the one-dimensional pair-source distribution in the longitudinal co-moving system (LCMS). The LCMS pair-separation vector can be expressed in lab-frame single-particle coordinates as
\begin{equation}
    \bs r_{LCMS} = \left(x_1{-}x_2,\;y_1{-}y_2,\;z_1{-}z_2{-}\frac{\beta(t_1{-}t_2)}{\sqrt{1{-}\beta^2}}\right), \textrm{ where } \beta = \frac{p_{z,1}{+}p_{z,2}}{E_1{+}E_2}.
\end{equation} 

Using this variable, one can construct the spatio-temporal distance distribution $D(\bs{r}_{LCMS},t)$. After angle- and time-integration we obtain the one-dimensional distance distribution as 
\begin{equation}
    D(r_{LCMS}) = \int d\Omega_{LCMS} dt D(\bs{r}_{LCMS},t), 
    \label{eq:drlcms}
\end{equation}
where we now suppress the $K$ dependence indicated in Eq.~\ref{eq:drk}. Note that the dependence on the lab-frame time-coordinate disappears after the time integral of Eq.~\ref{eq:drlcms}, since we only keep the dependence on $r_{LCMS}$, the final variable. In fact when analyzing the EPOS output, we only calculate the number of pairs in a given $r_{LCMS}$ bin, hence dependence on all other coordinates is naturally integrated out. When selecting pions we used the single-particle rapidity and transverse momentum requirements of $|\eta| < 1$ and $0.2 \textnormal{ GeV}/c~<~p_T~<~1.0~\textnormal{ GeV}/c$. For each individual event we constructed the $D(r_{LCMS})$ distribution for 5 different average transverse momentum $k_T$ classes in equal bins ranging from 0.2 to 0.4 GeV$/c$. Note that $k_T = 0.5\sqrt{K_x^2+K_y^2}$ is the transverse component of $K$ used in Eq.~\ref{eq:drk}. We chose this $k_T$ region to be around the peak of the pair $k_T$ distribution to have adequate statistics (number of pairs) in the individual $k_T$ bins. To investigate centrality dependence as well, we separated the measurements to the centrality classes of 0-5\%, 5-10\%, 10-20\%, 20-30\%. In total we used 63000 EPOS events, 10500 for the first two centrality classes and 21000 for the rest.

\begin{figure}
    \centering
    \includegraphics[width=0.6\textwidth]{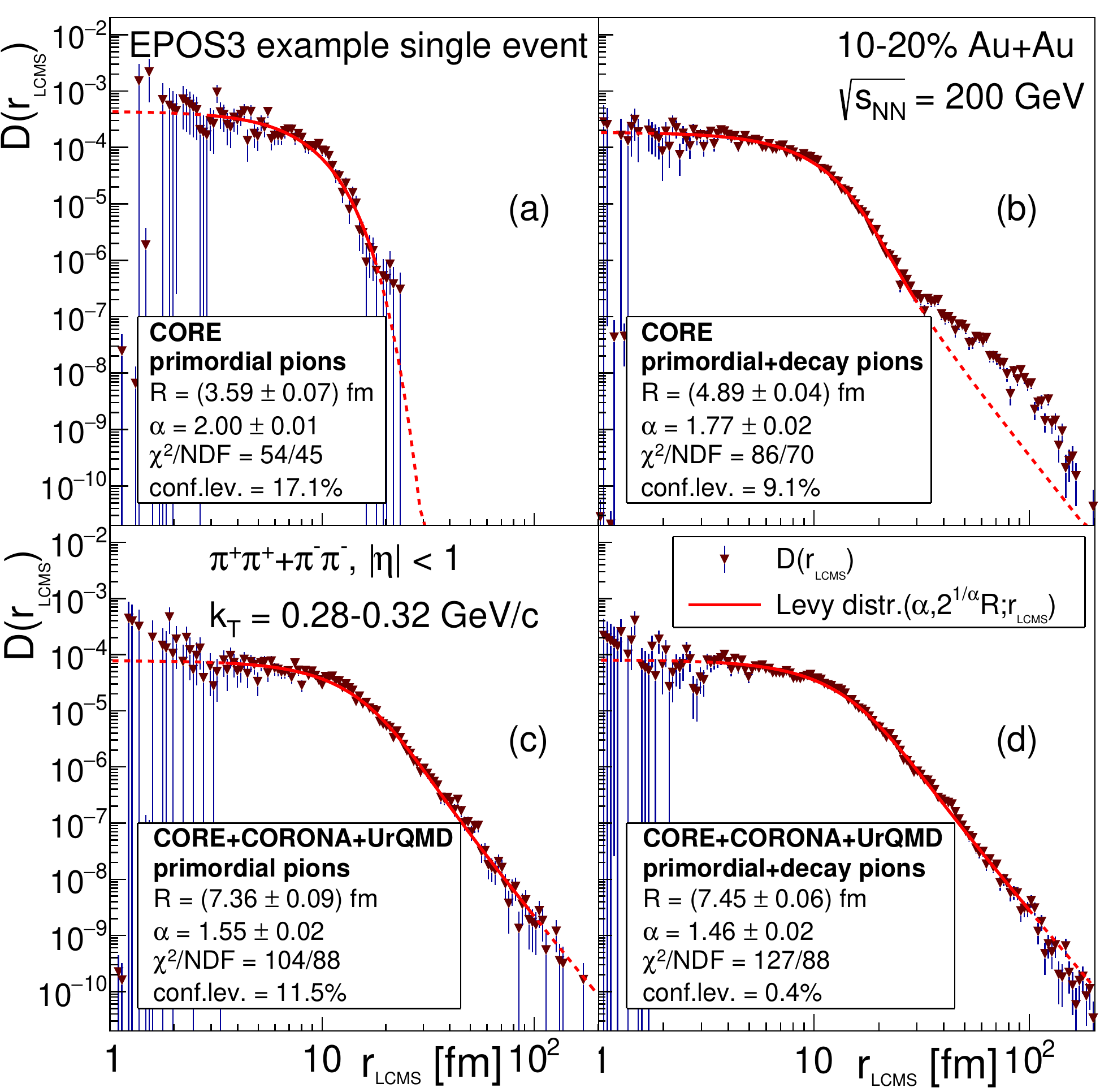}
    \caption{EPOS3 example single event fits for 10-20\% Au+Au collisions at $\sqrt{s_{NN}} = 200$ GeV. The average transverse momentum $k_T$ of the pion pairs is within the range of 0.28 GeV/$c$ to 0.32 GeV/$c$. The measured D($r_{LCMS}$) pion pair source distributions are denoted with downward-pointing triangles, and the fitted L\'evy distributions are plotted with a red line (continuous in the fit region, dashed outside). The four panels from left to right, top to bottom are \ref{c:c_prim} CORE with only primordial pions, \ref{c:c_all} CORE with primordial + decay pions, \ref{c:ccu_prim} CORE+CORONA+UrQMD with only primordial pions, and \ref{c:ccu_all} CORE+CORONA+UrQMD with primordial + decay pions.} 
    \label{f:levyfit}
\end{figure}

As mentioned before, EPOS has different stages of evolution including hydrodynamic expansion and hadronic rescattering. In order to identify the effect of the different stages on the shape of the pair-source distribution, as well as the contribution from the resonance decay products, we separated our investigation to four different cases as follows:
\begin{enumerate}[label=(\alph*), labelindent=\parindent, leftmargin=*]
    \item CORE with only primordial pions \label{c:c_prim}
    \item CORE with primordial + decay pions \label{c:c_all}
    \item CORE+CORONA+UrQMD with only primordial pions \label{c:ccu_prim}
    \item CORE+CORONA+UrQMD with primordial + decay pions, \label{c:ccu_all}
\end{enumerate}
where primordial pions include pions coming from the thermal medium, i.e., primordial pions are those that are not decay products. For each single event we fitted a L\'evy distribution to the constructed $D(r_{LCMS})$ distribution within the range of 2 fm to 100 fm. The fit was considered good if the confidence level calculated from the $\chi^2$ and $NDF$ values was greater than 0.1\%. An example of such fits for the four different cases can be seen on Figure~\ref{f:levyfit}. 

In case \ref{c:c_prim} we found that the events exhibit sharp cutoff features and mostly can be fitted well with a Gaussian. In case \ref{c:c_all} the inclusion of the decay product pions results in power-law like structures, appearing at different regions in $r_{LCMS}$. The shape of the events can be very different depending on the number (and origin) of decay pions in the sample. Due to the increased fluctuations and different event shapes the event-by-event extraction of the source parameters could not be done in the previous two cases. The fit settings would have to be fine-tuned for each event separately for this to work, which makes it impossible to do for thousands of events. In case \ref{c:ccu_prim} and \ref{c:ccu_all} however, distinct non-Gaussian structures (power-law tails) are present in all events, shapes can be described by L\'evy distributions in a statistically acceptable manner, and the extraction of the event-by-event source parameters is feasible. Furthermore, their distribution in the event sample can also be determined.

\begin{figure}
    \centering
    \includegraphics[width=0.65\textwidth]{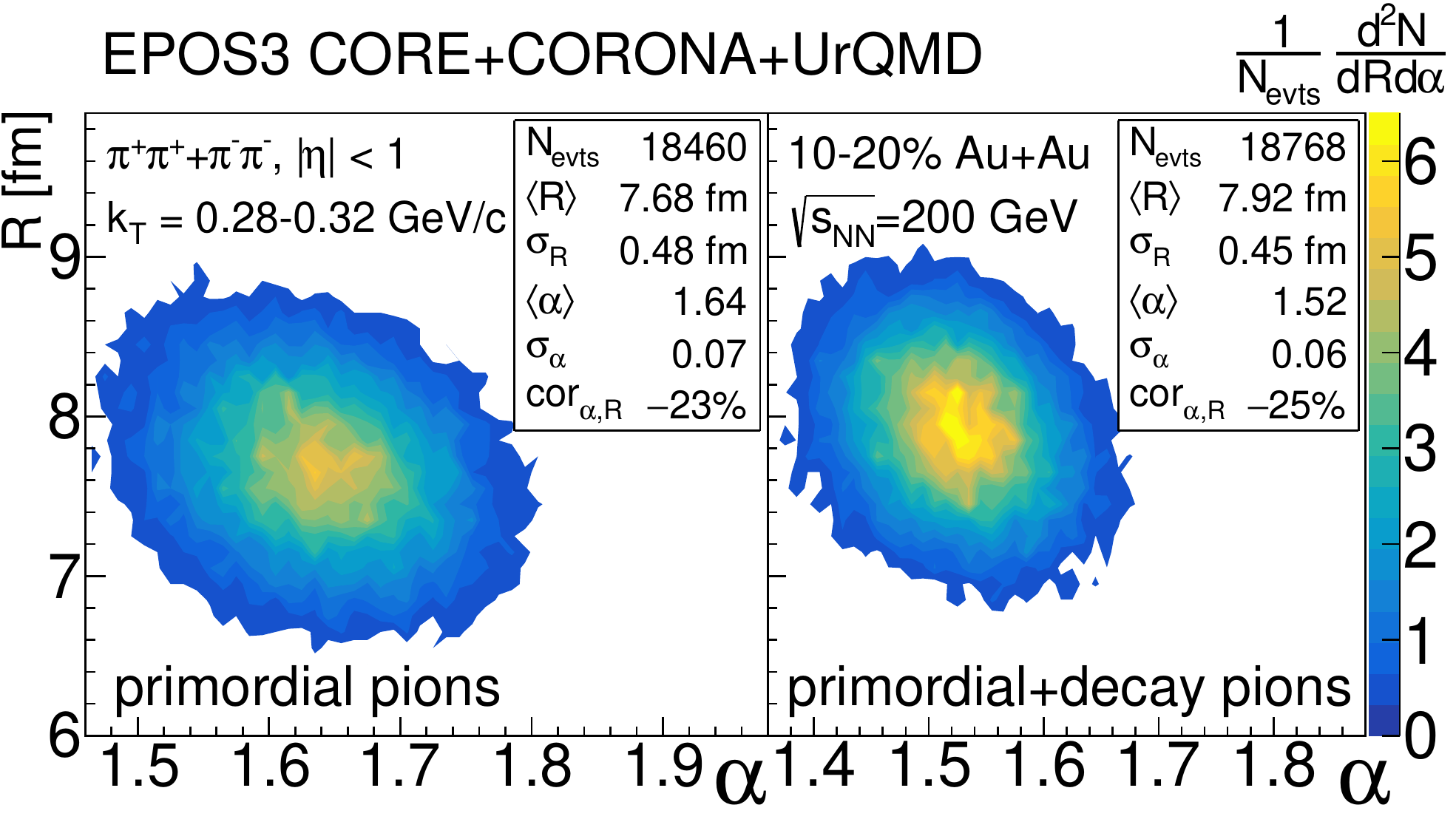}
    \caption{Example source parameter distributions for EPOS3 CORE+CORONA+UrQMD with only primordial pions (left) and with primordial and decay pions (right). The distribution was reconstructed from 10-20\% centrality $\sqrt{s_{NN}} = 200$ GeV Au+Au events for pion pairs with an average transverse momentum $k_T$ between 0.28 GeV/$c$ and 0.32 GeV/$c$.}
    \label{fig:Rvsalpha}
\end{figure}

An example for such a distribution can be seen on Figure \ref{fig:Rvsalpha}. In this example the distribution was reconstructed from fitting 21000 events, out of which  18460 fits were successful for case \ref{c:ccu_prim} and 18768 for case \ref{c:ccu_all} according to the confidence level criteria. Note that the distribution of source parameters was quite the same for non-acceptable fits as well, however, those do not necessarily represent the acquired source distributions, hence they were omitted from the further calculations. As mentioned before, we measured these two-dimensional $R$ vs. $\alpha$ distributions for 4 different centrality classes, and 5 different $k_T$ regions. From these we can extract the mean and standard deviation values, and investigate their centrality and $k_T$ dependence. There are multiple ways to determine the mean and standard deviation parameters -- on one hand, one could do normal distribution fits to the obtained 2D histograms, on the other hand one can simply calculate the first and second momenta of the distributions. We checked both and since the results were quite similar, for the sake of simplicity we chose the latter one.

Let us reiterate the point here that we analyzed individual EPOS events, and determined the $R$ and $\alpha$ parameters of the pair source distribution in those individual events. We saw L\'evy-shaped distributions when we included hadronic scattering, with slightly different $\alpha$ parameter values when decay pions were also included (besides primordial pions).

\section{Results and Discussion}
\label{s:results}

\begin{figure}[t]
    \centering
    \includegraphics[width=0.65\textwidth]{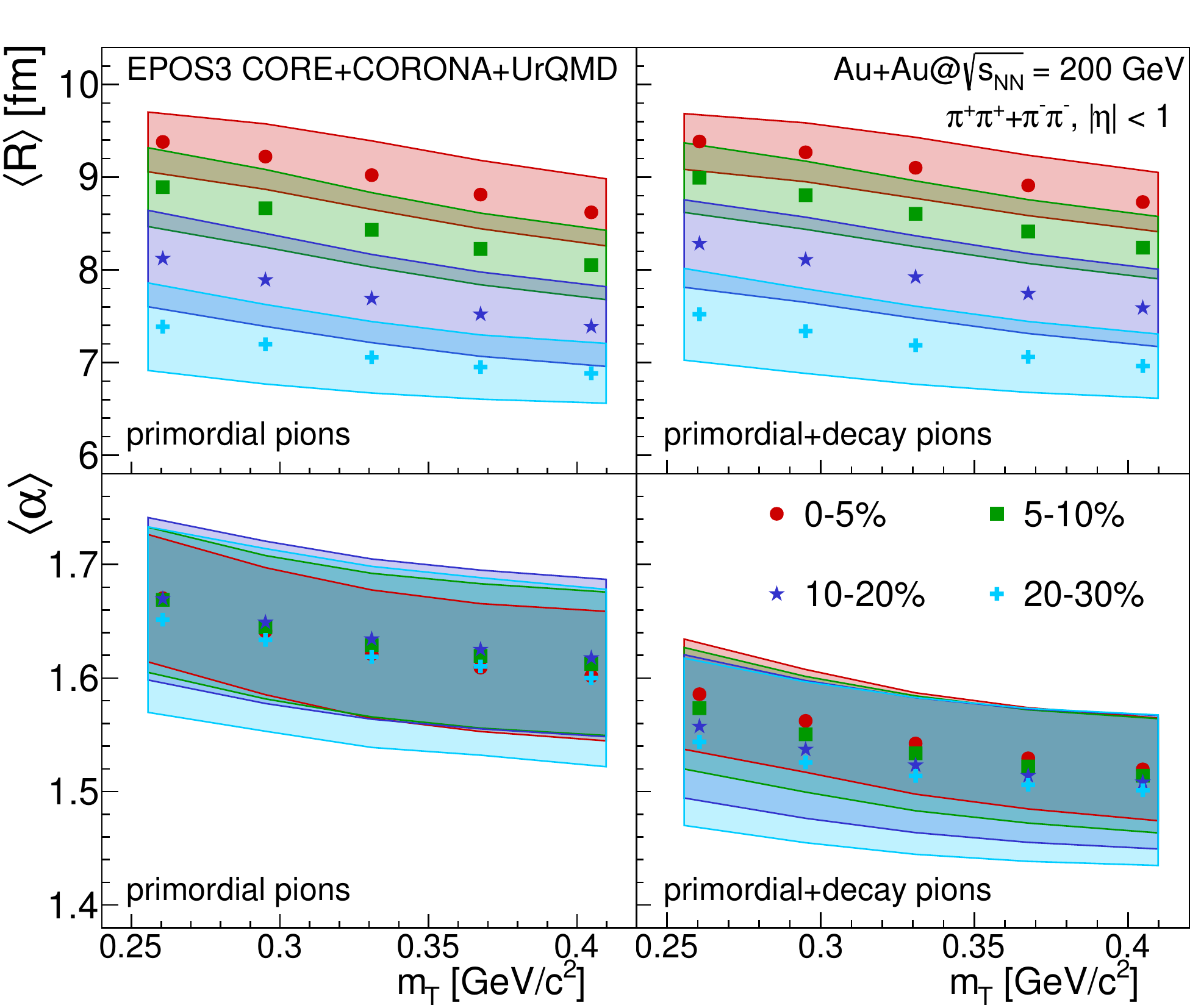}
    \caption{Mean $R$ and $\alpha$ values vs. average transverse mass $m_T$ for four different centrality classes (0-5\%, 5-10\%, 10-20\%, 20-30\%), in case of $\sqrt{s_{NN}} = 200$ GeV Au+Au collisions generated by EPOS3. The left column corresponds to the case of using primordial pions only, while the right column corresponds to the case of including both primordial and decay pions in the sample. The mean values are plotted with different filled markers. The corresponding colored boxes are representing the standard deviation values. The data from this Figure is listed in Table \ref{t:ccu} for both cases.}
    \label{f:RalphavsmT}
\end{figure}

As described above, we performed fits to individual EPOS events from the final stage of EPOS (CORE+CORONA+UrQMD), and investigated averages of the resulting $R$ and $\alpha$ parameters. We repeated this exercise for various centralities ($0-5\%$, $5-10\%$, $10-20\%$ and $20-30\%)$ and $k_T$ regions (5 equal bins from 0.2 to 0.4 GeV/$c$). We analyzed two cases separately: first the case of using only primordial pions, and then a case where both primordial and decay pions were included in the sample. Results for the mean $R$ and $\alpha$ values versus $m_T=\sqrt{m^2+k_T^2}$ are shown in Fig.~\ref{f:RalphavsmT}. 

One can observe a clear decreasing trend in $R$ with both $m_T$ and centrality, for both the case of primordial pions, as well as the case where decays were also included. This trend with $m_T$ is similar to the observed $R_{\rm Gauss}^2 \propto m_T$ trend observed universally across collision centrality, particle type, colliding energy, and colliding system size~\cite{Adler:2004rq,Afanasiev:2009ii}, even though it is based on Gaussian source radii. The decrease of $R$ with increasing centrality shows the relation of the Lévy-scale to the initial fireball size. One can also observe that $R$ is only weakly affected by the inclusion of decay pions; the values are slightly higher in the latter case. 

The Lévy-stability index $\alpha$ shows less prominent centrality dependence, although a small decrease for more peripheral events is visible in case we include decay pions as well. This feature, i.e. the centrality dependence of $\alpha$, was not yet investigated in experimental publications (with the exception of preliminary data in Ref.~\cite{Lokos:2018dqq}). Also, one may observe a weak decrease with $m_T$ similarly to what was observed in Ref.~\cite{Adare:2017vig}. Furthermore it is clearly visible that when decay pions are also included, the $\alpha$ parameter decreases. This is expected as decay pions produce an even stronger tail, creating a smaller $\alpha$ value.

Concrete values of $R$ can be compared to measured values from Ref.~\cite{Adare:2017vig}. There $R$ values of 7-8 fm were measured for the $0-30\%$ centrality class and the $m_T=0.25-0.45$ GeV/$c^2$ window. Our calculation yields similar values in the $10-20\%$ and $20-30\%$ centrality class. In Ref.~\cite{Adare:2017vig}, however, $\alpha$ values around 1.2 with a weak $m_T$ dependent decrease were found. The trend in $m_T$ is similar in our analysis, but the magnitude of the $\alpha$ values is somewhat different. The reasons for this could be multifold, they can range from the unavoidable event averaging present in the experiment to initial or final state effects not present in our simulations. The exploration of this difference is beyond the scope of present paper.

Note that the filled bands on both plots indicate the standard deviation of the $R$ and $\alpha$ distributions over the investigated event sample (let us remind the reader that we investigate and fit pair source distributions in individual events). The statistical uncertainty of these data points is basically negligible, due to the large number of events this average was performed over. This means that the trends observed in Fig.~\ref{f:RalphavsmT} and discussed above are true features of the EPOS event sample investigated in this paper.

\section{Summary and Conclusions}
\label{s:summary}
We investigated a sample of EPOS events individually, and using identical pion pairs we reconstructed the pair source function in every individual event. In the case when only primordial pions were analyzed before hadronic scattering, a Gaussian shape was observed. However, when decay pions were also included, already power-law like structures appeared, and after the inclusion of hadronic scatterings (via UrQMD), Lévy-shaped pair distributions arose in the individual events. It is hence clear that it is not the event averaging that creates the non-Gaussian features in the pair distributions (and the arising correlation functions in femtoscopical measurements).

Subsequently, using the final stage of EPOS events (CORE+CORONA+UrQMD) we analyzed the event sample mean of the event-by-event Lévy-scale $R$ and Lévy-index $\alpha$ values for the case of using only primordial pions and the case of including both primordial and decay pions. We observed clear trends as a function of $m_T$ and centrality for both cases. These observations show that in a realistic hydrodynamics-based simulation deviations from the Gaussian source shape appear when one includes hadronic scattering and decays. The values and trends of $R$ are compatible with experimentally measured values, although we did not perform a detailed data comparison here. The weak decrease of $\alpha$ with $m_T$ is similar to what was observed experimentally, but the values of $\alpha$ are somewhat larger than measured values.

In the future we plan to utilize similar techniques to explore the dependence of these results on particle species as well. Further investigations might also include expanding the analysis to multiple dimensions, different collision energies, as well as reconstructing femtoscopical correlation functions.

Finally let us note, that one of the important conclusions of our analysis is that the Lévy-shaped source assumption provides an acceptable description of the pion pair-source in EPOS3.


\authorcontributions{Conceptualization, M.Cs.; formal analysis, D.K.; resources, M.S.; software, D.K., M.S.; writing---original draft preparation, D.K.,M.Cs.,M.S.; writing---review and editing, M.Cs.,D.K.; visualization, D.K.; supervision, M.Cs. All authors 
have read and agreed to the published version of the manuscript.}

\funding{Our research has been partially supported by the Hungarian NKFIH Grants No. FK-123842, K-128713, K-138136, and 2019-2.1.11-T\'ET-2019-00080. D.K. was also supported by the ÚNKP-21-4 New National Excellence Program of the Hungarian Ministry for Innovation and Technology from the source of the National Research, Development and Innovation Fund. We also acknowledge the usage of computer cluster DWARF at Warsaw University of Technology supported by the Polish National Science Center (NCN) under Contracts No. UMO-2017/26/E/ST3/00428 and UMO-2017/27/B/ST2/02792. These studies were also funded by IDUB-POB-FWEiTE-1 project granted by Warsaw University of Technology under the program Excellence Initiative: Research University (ID-UB). M.S. also wishes to acknowledge the financial support of the German Academic Exchange Service (DAAD).}

\dataavailability{Data sharing is not applicable to this article.} 

\acknowledgments{The authors would like to thank H. Zbroszczyk, K. Werner and M. I. Nagy for useful discussions.}

\conflictsofinterest{The authors declare no conflict of interest.} 
\appendixtitles{no} 
\appendixstart
\appendix
\section{}
The data from Figure \ref{f:RalphavsmT}. is listed in Table \ref{t:ccu}.
\begin{specialtable}[H] 
\small
\setlength\tabcolsep{4pt}
\caption{Mean and standard deviation values extracted from the source parameter distributions measured in EPOS3 Au+Au collisions at $\sqrt{s_{NN}} = 200$ GeV. For the average transverse momentum $k_T$ the range is indicated, while for the average transverse mass $m_T$ the central value is given. The correlation coefficients and the covariance values are also indicated. \label{t:ccu}}
\begin{tabular}{ccccccccc}
\toprule
\multicolumn{9}{c}{primordial pions}\\
centr. & $k_T$ [GeV/$c$] & $m_T$ [GeV/$c^2$] & $\boldsymbol{\langle R \rangle}$ [fm]    &   $\boldsymbol{\sigma_R}$ [fm]    &   $\boldsymbol{\langle \alpha \rangle}$    &   $\boldsymbol{\sigma_\alpha}$    & cor($R,\alpha$)  & cov($R,\alpha$) [fm]\\
\midrule
\multirow{5}{*}{0-5\%} & 0.20-0.24 & 0.261 & 9.381 & 0.322 & 1.670 & 0.056 & -0.225 & -0.004\\
 & 0.24-0.28 & 0.295 & 9.222 & 0.353 & 1.641 & 0.056 & -0.273 & -0.005\\
 & 0.28-0.32 & 0.331 & 9.021 & 0.370 & 1.621 & 0.056 & -0.305 & -0.006\\
 & 0.32-0.36 & 0.368 & 8.811 & 0.368 & 1.610 & 0.056 & -0.319 & -0.007\\
 & 0.36-0.40 & 0.405 & 8.619 & 0.361 & 1.603 & 0.056 & -0.324 & -0.007\\
\midrule
\multirow{5}{*}{5-10\%} & 0.20-0.24 & 0.261 & 8.892 & 0.425 & 1.669 & 0.064 & -0.328 & -0.009\\
 & 0.24-0.28 & 0.295 & 8.663 & 0.419 & 1.645 & 0.063 & -0.326 & -0.009\\
 & 0.28-0.32 & 0.331 & 8.431 & 0.401 & 1.630 & 0.062 & -0.329 & -0.008\\
 & 0.32-0.36 & 0.368 & 8.223 & 0.385 & 1.621 & 0.062 & -0.332 & -0.008\\
 & 0.36-0.40 & 0.405 & 8.049 & 0.372 & 1.614 & 0.061 & -0.316 & -0.007\\
\midrule
\multirow{5}{*}{10-20\%} & 0.20-0.24 & 0.261 & 8.121 & 0.521 & 1.671 & 0.071 & -0.229 & -0.008\\
 & 0.24-0.28 & 0.295 & 7.888 & 0.504 & 1.650 & 0.070 & -0.235 & -0.008\\
 & 0.28-0.32 & 0.331 & 7.683 & 0.480 & 1.636 & 0.068 & -0.230 & -0.008\\
 & 0.32-0.36 & 0.368 & 7.510 & 0.463 & 1.627 & 0.067 & -0.224 & -0.007\\
 & 0.36-0.40 & 0.405 & 7.370 & 0.444 & 1.620 & 0.066 & -0.208 & -0.006\\
\midrule
\multirow{5}{*}{20-30\%} & 0.20-0.24 & 0.261 & 7.350 & 0.502 & 1.655 & 0.078 & -0.142 & -0.006\\
 & 0.24-0.28 & 0.295 & 7.135 & 0.471 & 1.639 & 0.076 & -0.128 & -0.005\\
 & 0.28-0.32 & 0.331 & 6.971 & 0.438 & 1.627 & 0.073 & -0.137 & -0.004\\
 & 0.32-0.36 & 0.368 & 6.844 & 0.407 & 1.619 & 0.072 & -0.119 & -0.003\\
 & 0.36-0.40 & 0.405 & 6.749 & 0.390 & 1.610 & 0.071 & -0.114 & -0.003\\
\bottomrule
\multicolumn{9}{c}{primordial + decay pions}\\
centr. & $k_T$ [GeV/$c$] & $m_T$ [GeV/$c^2$] & $\boldsymbol{\langle R \rangle}$ [fm]    &   $\boldsymbol{\sigma_R}$ [fm]    &   $\boldsymbol{\langle \alpha \rangle}$    &   $\boldsymbol{\sigma_\alpha}$    & cor($R,\alpha$)  & cov($R,\alpha$) [fm]\\
\midrule
\multirow{5}{*}{0-5\%} & 0.20-0.24 & 0.261 & 9.385 & 0.301 & 1.586 & 0.049 & -0.287 & -0.004\\
 & 0.24-0.28 & 0.295 & 9.269 & 0.318 & 1.562 & 0.045 & -0.224 & -0.003\\
 & 0.28-0.32 & 0.331 & 9.101 & 0.329 & 1.542 & 0.045 & -0.265 & -0.004\\
 & 0.32-0.36 & 0.368 & 8.911 & 0.326 & 1.529 & 0.045 & -0.277 & -0.004\\
 & 0.36-0.40 & 0.405 & 8.732 & 0.320 & 1.520 & 0.045 & -0.309 & -0.004\\
\midrule
\multirow{5}{*}{5-10\%} & 0.20-0.24 & 0.261 & 8.995 & 0.375 & 1.573 & 0.054 & -0.311 & -0.006\\
 & 0.24-0.28 & 0.295 & 8.805 & 0.368 & 1.550 & 0.051 & -0.292 & -0.005\\
 & 0.28-0.32 & 0.331 & 8.605 & 0.355 & 1.534 & 0.051 & -0.305 & -0.005\\
 & 0.32-0.36 & 0.368 & 8.413 & 0.344 & 1.522 & 0.050 & -0.309 & -0.005\\
 & 0.36-0.40 & 0.405 & 8.240 & 0.336 & 1.514 & 0.050 & -0.313 & -0.005\\
\midrule
\multirow{5}{*}{10-20\%} & 0.20-0.24 & 0.261 & 8.284 & 0.472 & 1.557 & 0.063 & -0.236 & -0.007\\
 & 0.24-0.28 & 0.295 & 8.109 & 0.460 & 1.537 & 0.061 & -0.237 & -0.007\\
 & 0.28-0.32 & 0.331 & 7.922 & 0.446 & 1.523 & 0.059 & -0.250 & -0.007\\
 & 0.32-0.36 & 0.368 & 7.744 & 0.433 & 1.514 & 0.059 & -0.253 & -0.006\\
 & 0.36-0.40 & 0.405 & 7.586 & 0.421 & 1.508 & 0.059 & -0.263 & -0.007\\
\midrule
\multirow{5}{*}{20-30\%} & 0.20-0.24 & 0.261 & 7.500 & 0.517 & 1.544 & 0.074 & -0.252 & -0.010\\
 & 0.24-0.28 & 0.295 & 7.305 & 0.489 & 1.527 & 0.071 & -0.256 & -0.009\\
 & 0.28-0.32 & 0.331 & 7.132 & 0.465 & 1.516 & 0.070 & -0.261 & -0.008\\
 & 0.32-0.36 & 0.368 & 6.982 & 0.437 & 1.509 & 0.068 & -0.256 & -0.008\\
 & 0.36-0.40 & 0.405 & 6.856 & 0.411 & 1.505 & 0.067 & -0.242 & -0.007\\
\bottomrule
\end{tabular}
\end{specialtable}
\end{paracol}
\newpage
\reftitle{References}

\externalbibliography{yes}
\bibliography{main}

\end{document}